\documentclass[a4paper,11pt,showpacs]{revtex4-1}
\usepackage{graphicx}
\usepackage{amssymb}
\usepackage{amsmath}
\usepackage{bm}
\begin{document}

\title{Unified parameter for localization in isotope-selective rotational excitation of diatomic molecules using a train of optical pulses}

\author{Leo Matsuoka}
\email[]{leo-matsuoka@hiroshima-u.ac.jp}

\affiliation{Graduate School of Engineering, Hiroshima University, Kagamiyama, Higashi-Hiroshima, 739-8527,Japan}

\date{\today}

\begin{abstract}

We obtained a simple theoretical unified parameter for the characterization of rotational population propagation of diatomic molecules in a periodic train of resonant optical pulses.
The parameter comprises the peak intensity and interval between the pulses, and the level energies of the initial and final rotational states of the molecule.
Using the unified parameter, we can predict the upper and lower boundaries of probability localization on the rotational level network, including the effect of centrifugal distortion.
The unified parameter was tentatively derived from an analytical expression obtained by performing rotating-wave approximation and spectral decomposition of the time-dependent Schr\"{o}dinger equation under an assumption of time-order invariance.
The validity of the parameter was confirmed by comparison with numerical simulations for isotope-selective rotational excitation of KCl molecules.
\end{abstract}

\thanks{Submitted to Phys. Rev. A}
\maketitle

\section{Introduction}
The rotation of small molecules has been studied from the viewpoint of non-adiabatic molecular alignment \cite{al01,al02,al03} or molecular orientation \cite{ori01,ori02,ori03,ori04,ori05,ori06,ori07}.
Non-adiabatic molecular alignment has been experimentally realized using non-resonant broadband optical pulses \cite{alex01,alex02}.
Recently, a train of pulses, whose repetition intervals are synchronized with the classical rotational period of the molecules, have been shown to be efficient for the rotational excitation of diatomic molecules \cite{pt01,pt02}.
At present, because of the development of terahertz-wave technology \cite{thz01,thz02,thz03}, molecular orientation by rotationally resonant terahertz pulses has also been studied \cite{oriex01,oriex02,oriex03,oriex04}.
The elemental technology employed to create a train of pulses in the terahertz region has already been reported \cite{thzpt01}.
An experiment involving molecular orientation using a train of rotationally resonant pulses will be performed in the near future.

Using a train of optical pulses, we can impart the isotope selectivity in rotational excitation, because the classical rotational period depends on the mass of the molecules \cite{iso01,iso02,iso03,iso04,iso05}.
The isotope selective population displacement of rotationally resonant terahertz pulses by the train is particularly considered to be practical \cite{iso_yo,iso_ichi,global}.
Because the consecutive rotational population transfer by the train of pulses simultaneously starts from all the rotational states in the molecule,
the principle of isotope selection still works well
even if the initial rotational energies of the molecular ensemble are thermally distributed, specifically at high temperatures.
However, there are still many problems with the practical application of isotope separation of optical pulses in the terahertz region by the train.
Further theoretical studies and development of elemental technologies are required.

\begin{figure}[htbp]
  \includegraphics[keepaspectratio,height=6.0cm]{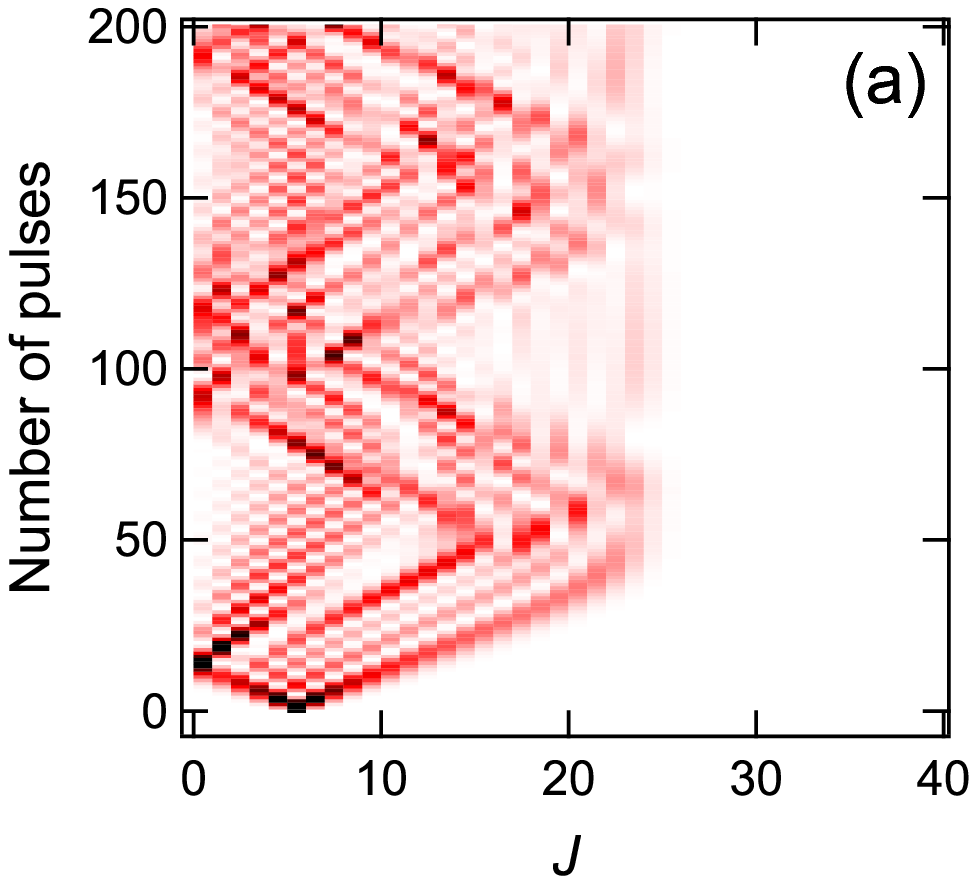}
  \includegraphics[keepaspectratio,height=6.0cm]{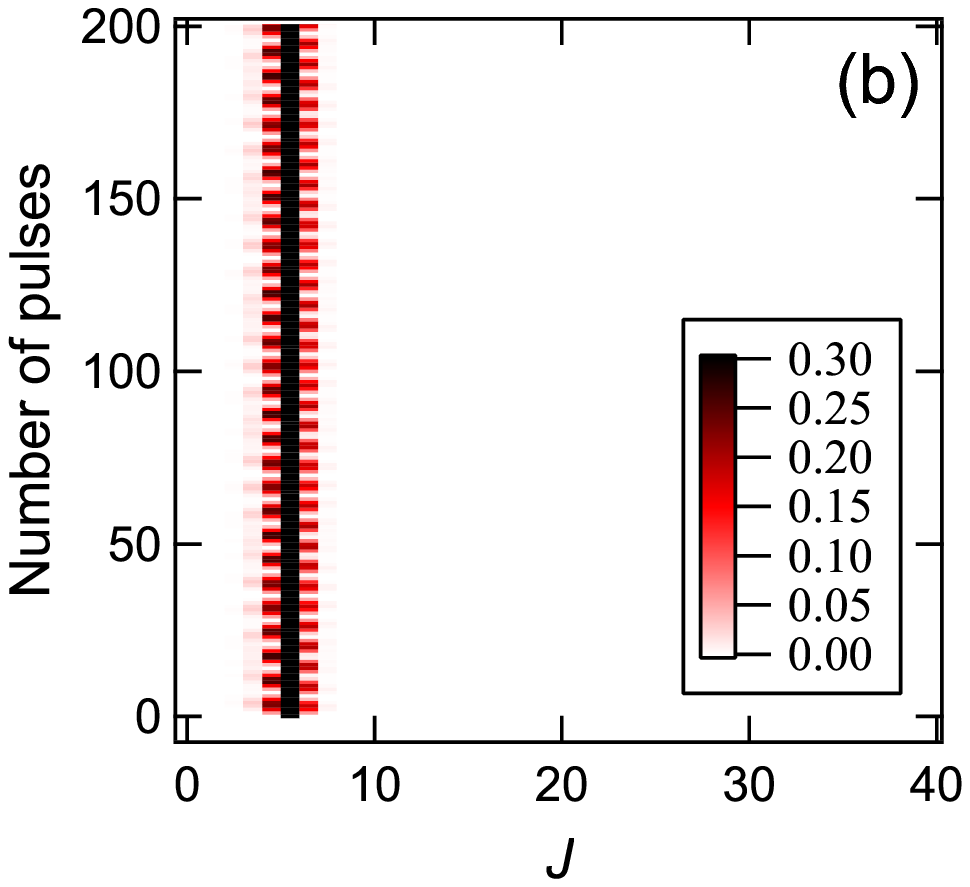}
  \caption{\label{fig_exp}Examples of numerical simulations of the evolution of rotational distribution for $J=5$ of $^{39}$K$^{37}$Cl (a) and $^{39}$K$^{35}$Cl (b), by irradiating the resonant pulse train synchronized with rotational period of $^{39}$K$^{37}$Cl. The color scale is the same in both (a) and (b).}
\end{figure}

In real diatomic molecules, the consecutive rotational excitation by the pulse train has a certain upper boundary even if the interval of pulses is synchronized with the rotational period \cite{floss01,floss02,floss03}.
For a single pulse excitation, the limitation ordinarily comes from the spectral bandwidth of the pulse. 
For a multi-pulse scheme, two additional mechanisms induce the localization of rotational distribution;
centrifugal distortion and mismatching of the pulse interval.
Figure 1 shows an example of the time-evolution of the rotational probability distribution
of diatomic molecules in a periodic train of pulses.
For the isotope in a synchronized pulse train (Fig. \ref{fig_exp}(a)), the wave packet of rotational probability is reflected at a certain limit, and repeatedly excited and de-excited inside the region.
This phenomenon is localization by centrifugal distortion, called Bloch oscillation by Flo{\ss} {\it et al} \cite{floss03}.
For another isotope (Fig. \ref{fig_exp}(b)), the population distribution was localized between $J=4$ and $J=6$.
This is localization by interval mismatching, called Anderson localization by Flo{\ss} {\it et al} \cite{floss03}.

We can intuitively understand the localization in the train of optical pulses by regarding the interaction as the resonant transition induced by the optical frequency comb.
The train of optical pulses whose interval is $T_{\mbox{\scriptsize{p}}}$ composes the optical frequency comb whose frequency peaks are evenly spaced by $1/T_{\mbox{\scriptsize{p}}}$.
Ignoring the centrifugal distortion,
the one-photon pure rotational transition frequencies of the diatomic molecules are also evenly spaced.
Isotope selectivity in rotational excitation originates from the agreement of the peak frequencies between the optical frequency comb and the molecular rotational transitions.
Mismatching of $T_{\mbox{\scriptsize{p}}}$ removes the comb's peak frequencies from the resonance, and the consecutive resonant transitions are immediately prevented.
The effect of centrifugal distortion can be ignored around the ground state; however, this effect becomes critical in the rotationally excited states.
The centrifugal distortion shifts the transition frequency in proportion to $(J+1)^3$.
Such a shift cannot be compensated by the optical frequency comb as long as only Fourier-transform-limited pulses are used.

A quantitative evaluation of the region of localization of rotational distribution is important in the planning of experiments.
This evaluation can be conducted by calculating the evolution of the time-dependent Schr\"{o}dinger equation system; however, the dependence of the region of localization on the intensity of the pulses is odd and quite unclear.
Therefore, to obtain the upper and lower boundaries of localization, it is necessary to perform numerical simulations for each case.

In this study, we propose an evaluation method for the localization of rotational distribution without performing numerical simulations.
We considered an evenly spaced Fourier-transform-limited pulse train in the terahertz region.
We create a mathematical model using approximations to the time-dependent Schr\"{o}dinger equation system for diatomic molecules, and tentatively derive an analytical solution by partly neglecting precision.
We extract a critical unified parameter from the analytical solution and validate it through comparison with the result of numerical simulations.
In Sec. 2, we begin with an introduction to the targeted physical system and theoretically derive the unified parameter.
In Sec. 3, the method used for the numerical simulations is provided.
In Sec. 4, we show the results of the numerical simulations to be compared with the theoretical predictions. 
In Sec. 5, we discuss the results and meaning of the theoretical analysis.
In Sec. 6, we provide the conclusions to this study.
Atomic units $\hbar=m_e=e=1$, $h=2\pi$ are used unless explicitly stated.

\section{Mathematical modeling}
\subsection{Physical system}
We consider a one-dimensional network comprising a series of rotational states in diatomic molecules, which are connected by resonant optical transitions.
In the interaction picture, the time-dependent Schr\"{o}dinger equation of the diatomic molecules in a linearly polarized electric field is given by
\begin{eqnarray}
\label{eq_sch_ori}
i\frac{d}{dt}C_J(t)= - \varepsilon (t)[&\mu_{J-1}& \cdot \exp \{(E_J-E_{J-1})\cdot it \} \cdot C_{J-1}(t) \nonumber \\ 
 + &\mu_{J}& \cdot \exp \{(E_J-E_{J+1})\cdot it \}\cdot C_{J+1}(t) ],
\end{eqnarray}
where $C_J(t)$ is the complex amplitude of the rotational states $J$ at time $t$, $\varepsilon(t)$ is the electric field, $\mu_J$ is the transition dipole moment from $J$ to $J+1$, and $E_J$ is the rotational energy of state $J$.
In the diatomic molecules, the rotational energy $E_J$ and the transition frequencies from $J$ to $J+1$ are given by
\begin{eqnarray}
\label{eq_E_j}
E_J &=& 2\pi B_{\mbox{\scriptsize{M}}}J(J+1) - 2\pi D_{v}J^2(J+1)^2, \\
\label{eq_nu_j}
\nu_J &=& (E_{J+1}-E_J)/2\pi, \nonumber \\
      &=& 2B_{\mbox{\scriptsize{M}}}(J+1) - 4D_{v}(J+1)^3,
\end{eqnarray}
where $B_{\mbox{\scriptsize{M}}}$ and $D_{v}$ are the molecular spectroscopic constants for a rotational series.
Generally, the ratio of these constants is very small ($D_{v}/B_{\mbox{\scriptsize{M}}}\cong 10^{-5}-10^{-7}$).
The classical rotational period of the diatomic molecules is defined by
\begin{equation}
\label{eq_t_mol}
T_{\mbox{\scriptsize{M}}}=\frac{1}{2B_{\mbox{\scriptsize{M}}}},
\end{equation}
ignoring centrifugal distortion.
The molecular transition dipole moment is expressed as
\begin{equation}
\label{eq_mu_j}
\mu_J = \mu_0\sqrt{\frac{(J+1)^2-M^2}{(2J+1)(2J+3)}},
\end{equation}
where $\mu_0$ is the permanent dipole moment,
and $M$ is the magnetic quantum number of the rotational series.
Because the electric field is supposed to be polarized linearly,
we need to consider only one $M$ in each series.
In our mathematical treatment, we approximate $\mu_J$ as
\begin{equation}
\label{eq_mu_half}
\mu_J \approx \frac{\mu_0}{2}.
\end{equation}
This approximation provides an almost exact solution for $M=0$,
and can be used for small $M$ in absolute values.
Using the expressions given here, eq.(\ref{eq_sch_ori}) can be rewritten as
\begin{eqnarray}
\label{eq_sch}
i\frac{d}{dt}C_J(t)= - \frac{\varepsilon (t)\mu_0}{2}[\exp \{2\pi \nu_{J-1}\cdot it \} \cdot C_{J-1}(t) \nonumber \\ 
 + \exp \{-2\pi \nu_{J}\cdot it \}\cdot C_{J+1}(t) ].
\end{eqnarray}

\begin{figure}[htbp]
  \includegraphics[keepaspectratio,width=8cm]{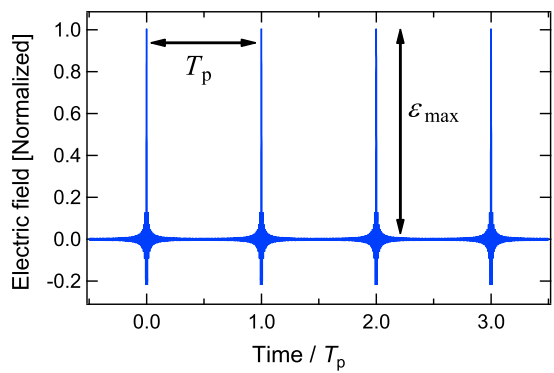}
  \caption{\label{fig_e_field}A typical time-profile of a train of optical pulses, calculated using eq.(\ref{eq_e_field}) with N=100.}
\end{figure}

We define the train of optical pulses using a multiple cosine electric field as
\begin{equation}
\label{eq_e_field}
\varepsilon (t) = \frac{\gamma B_{\mbox{\scriptsize{f}}}}{\mu_0} \left[ 1+2\sum_{j=0}^{N-1}\cos \{2\pi \cdot 2 B_{\mbox{\scriptsize{f}}}(j+1)t\}  \right],
\end{equation}
where $\gamma$ is the scaled intensity of the electric field,
$B_{\mbox{\scriptsize{f}}}$ is the frequency constant that corresponds to $B_{\mbox{\scriptsize{M}}}$ of the molecules,
and $N$ is the maximum $J$ state to be connected by optical transitions.
Under a rough estimation, the value of $\gamma$ corresponds to increments of $J$ in a single pulse interaction.
An example of the time-profile of the train of pulses is shown in Fig.\ref{fig_e_field}.
The interval between the pulses and the maximum amplitude of a pulse are given by \cite{blumel}
\begin{eqnarray}
\label{eq_t_pulse}
T_{\mbox{\scriptsize{p}}} &=& \frac{1}{2B_{\mbox{\scriptsize{f}}}},\\
\label{eq_e_max}
\varepsilon_{\mbox{\scriptsize{max}}} &=& \frac{\gamma B_{\mbox{\scriptsize{f}}}}{\mu_0}(2N+1).
\end{eqnarray}

\subsection{Rotating-wave approximation}
From here, we perform a mathematical treatment to obtain an analytical expression for the rotational probability distribution.
First, we apply the rotating-wave approximation,
in which only the slowest oscillating terms are used and other quickly oscillating terms are ignored.
We previously used this approximation to find the analogy between consecutive rotational excitation in a pulse train and continuous-time quantum walk (CTQW) \cite{JKPS}.
Under a fully resonant condition ($T_{\mbox{\scriptsize{p}}}=T_{\mbox{\scriptsize{M}}}, D_v=0$), 
the evolution of the rotational distribution was observed to agree with the evolution of the CTQW. 
In our numerical calculations, under the conditions in which localization occurs, the rotating-wave approximation derives almost exact numerical results for a relatively weak electric field ($\gamma<10$).

Substituting eq.(\ref{eq_e_field}) into eq.(\ref{eq_sch}), and applying the rotating-wave approximation, we obtain the Schr\"{o}dinger equation as
\begin{eqnarray}
\label{eq_sch_ra}
i\frac{d}{dt}C_J(t)= - \frac{\gamma \cdot B_{\mbox{\scriptsize{f}}}}{2}[&\exp& \{(\nu_{J-1}-4\pi B_{\mbox{\scriptsize{f}}}J)\cdot it \} \cdot C_{J-1}(t) \nonumber \\ 
 + &\exp& \{(-\nu_{J+1}+4\pi B_{\mbox{\scriptsize{f}}}(J+1))\cdot it \}\cdot C_{J+1}(t) ].
\end{eqnarray}
Without the approximation, eq. (\ref{eq_sch_ra}) involves many terms that oscillate with various frequencies in the exponential.
The slowest oscillating terms remain in the exponential by considering the probable experimental conditions, that is, $B_{\mbox{\scriptsize{f}}} \cong B_{\mbox{\scriptsize{M}}}$ and $D_v \ll B_{\mbox{\scriptsize{f}}}$.

\subsection{Decomposition of interaction matrix}
Using matrix representation, the interaction Hamiltonian of eq. (\ref{eq_sch_ra}) is given by,
\begin{equation}
  \label{eq_sch_array}
  \mbox{\boldmath{$\hat{H}$}}(t)=-\frac{\gamma \cdot B_{\mbox{\scriptsize{f}}}}{2}\left[
  \begin{array}{ccccc}
    0 & \mbox{e}^{-\alpha_0it} & & & 0 \\
    \mbox{e}^{\alpha_0it} & 0 & \mbox{e}^{-\alpha_1it} & & \\
       & \mbox{e}^{\alpha_1it} & \ddots & \ddots &  \\
      & & \ddots & 0 & \mbox{e}^{-\alpha_{N-1}it} \\
    0 &  &  & \mbox{e}^{\alpha_{N-1}it} & 0 \\
  \end{array}
  \right],
\end{equation}
where $\alpha_J$ is defined by
\begin{equation}
\label{eq_alpha}
\alpha_J=2\pi \{2\Delta B(J+1)-4D_v(J+1)^3\},
\end{equation}
where $\Delta B=B_{\mbox{\scriptsize{M}}}-B_{\mbox{\scriptsize{f}}}$.
The interaction matrix eq.(\ref{eq_sch_array}) is a time-dependent tridiagonal matrix, and is actually inconvenient for analytical treatment.
We introduce $\beta_J$ to decompose this matrix as
\begin{eqnarray}
\label{eq_beta}
\beta_J &=& 2 \pi \{\Delta BJ(J+1)- D_vJ^2(J+1)^2\},\\
 & = & E_J-2\pi B_{\mbox{\scriptsize{f}}} J(J+1),\\
 & = & E_J-\frac{\pi J(J+1)}{T_{\mbox{\scriptsize{p}}}}.
\end{eqnarray}
$\alpha_J$ is expressed using $\beta_J$ as
\begin{equation}
\label{eq_beta_alpha}
\alpha_J=\beta_{J+1}-\beta_J.
\end{equation}
Using $\beta_J$, eq. (\ref{eq_sch_array}) can be decomposed as
\begin{equation}
\label{eq_h_deco1}
\mbox{\boldmath{$\hat{H}$}}(t)=-\frac{\gamma \cdot B_{\mbox{\scriptsize{f}}}}{2}\mbox{\boldmath{$T$}}(t)\mbox{\boldmath{$V$}}\mbox{\boldmath{$\tilde{T}$}}(t), \\
\end{equation}
where
\begin{eqnarray}
\label{eq_T}
\mbox{\boldmath{$T$}}(t) & = & \left[
  \begin{array}{cccc}
    \mbox{e}^{\beta_0it} & 0 & \cdots & 0 \\
     0 & \mbox{e}^{\beta_1it} & \cdots & 0 \\
     \vdots & \vdots & \ddots & 0\\
     0 & 0 & 0 & \mbox{e}^{\beta_{N}it}  \\
  \end{array}
  \right], \\
\label{eq_V}
\mbox{\boldmath{$V$}} &=&\left[
  \begin{array}{ccccc}
    0 & 1 & 0 & \cdots & 0 \\
    1 & 0 & 1 & \cdots & 0 \\
    0 & 1 & 0 & \cdots & 0  \\
    \vdots & \vdots & \vdots & \ddots & 1 \\
    0 & 0 &  0 & 1 & 0 \\
  \end{array}
  \right],
\end{eqnarray}
where $\mbox{\boldmath{$\tilde{T}$}}(t)$ is the conjugate matrix of $\mbox{\boldmath{${T}$}}(t)$, namely, $\mbox{\boldmath{$\tilde{T}$}}(t)\cdot \mbox{\boldmath{$T$}}(t)=\mbox{\boldmath{$I$}}$, where $\mbox{\boldmath{$I$}}$ is the identity matrix.
Up until this point, the interaction matrix $\mbox{\boldmath{$\hat{H}$}}(t)$ was decomposed into time-dependent diagonal matrices and a time-independent tridiagonal matrix.

The matrix in eq.(\ref{eq_h_deco1}) can be decomposed again.
Using the method called spectral decomposition,
eq.(\ref{eq_V}) can be decomposed as
\begin{equation}
\label{eq_h_deco2}
\mbox{\boldmath{$V$}}=\mbox{\boldmath{$Y$}}\mbox{\boldmath{$\Lambda$}}\mbox{\boldmath{$\tilde{Y}$}}, \\
\end{equation}
where $\mbox{\boldmath{$\Lambda$}}$ is a diagonal matrix given by
\begin{eqnarray}
\label{eq_lambda}
\mbox{\boldmath{$\Lambda$}} &=& \left[
  \begin{array}{cccc}
    \lambda_0 & 0 & \cdots & 0 \\
     0 & \lambda_1 & \cdots & 0 \\
     \vdots & \vdots & \ddots & 0\\
     0 & 0 & 0 & \lambda_N  \\
  \end{array}
  \right],\\
\lambda_k &=& 2 \cos \left(\frac{(k+1)\pi}{N+1}\right), (0\leq k \leq N),
\end{eqnarray}
and $\mbox{\boldmath{$Y$}}$ is a unitary matrix given by
\begin{eqnarray}
\label{eq_Y}
\mbox{\boldmath{$Y$}} &=&\left[
  \begin{array}{cccc}
    y_{00} & y_{10} & \cdots & y_{N0} \\
    y_{01} & y_{11} & \cdots & y_{N1} \\
    \vdots & \vdots & \ddots & \vdots \\
    y_{0N} & y_{1N} & \cdots & y_{NN} \\
  \end{array}
  \right], \\
y_{jk} &=& \sqrt{\frac{2}{N+1}}\sin \left(\frac{(j+1)(k+1)\pi}{N+1}\right).
\end{eqnarray}
From the result above, the interaction matrix can be given as
\begin{eqnarray}
\label{eq_h_deco}
\mbox{\boldmath{$\hat{H}$}}(t) & =& -\frac{\gamma \cdot B_f}{2}\mbox{\boldmath{$T$}}(t)\mbox{\boldmath{$Y$}}\mbox{\boldmath{$\Lambda$}}\mbox{\boldmath{$\tilde{Y}$}}\mbox{\boldmath{$\tilde{T}$}}(t), % \\
%\left[\mbox{\boldmath{$\hat{H}$}}(t)\right]^n & =& \left(-\frac{\gamma \cdot B_f}{2}\right)^n \mbox{\boldmath{$T$}}(t)\mbox{\boldmath{$Y$}}\mbox{\boldmath{$\Lambda$}}^n\mbox{\boldmath{$\tilde{Y}$}}\mbox{\boldmath{$\tilde{T}$}}(t).
\end{eqnarray}

\subsection{Analytical solution with an assumption of time-order invariance}
From here, we derive an analytical solution for the time-evolution of the rotational probability amplitude.
Generally, for the case of matrices such as eq.(\ref{eq_sch_array}), 
we have to consider the time-ordered product,
which is the integral of the interaction Hamiltonian along all possible quantum paths.
The exact solution is given by the formula as 
\begin{equation}
\label{eq_integrate_exact}
\mbox{\boldmath{$C$}}(t)=\mbox{\boldmath{$C$}}(t_0)+\sum_{n=1}^{\infty}(-i)^n\int_{t_0}^t\!\! dt_{n-1}\int_{t_0}^{t_{n-1}} \!\! dt_{n-2}\! \cdots \!\!\int_{t_0}^{t_1}\!\!dt_0 \mbox{\boldmath{$\hat{H}$}}(t_{n-1})\mbox{\boldmath{$\hat{H}$}}(t_{n-2})\cdots \mbox{\boldmath{$\hat{H}$}}(t_{0})\mbox{\boldmath{$C$}}(t_0),
\end{equation}
where $\mbox{\boldmath{$C$}}(t)$ is a vector representation of $C_J(t)$.
However, in our case, because many quantum states should be considered, infinite quantum paths must be examined.
Such a mathematically exact treatment is impossible for the physical model presented.
Here we derive a tentative expression for the probability amplitude by ignoring mathematical precision.
We assume the time-order invariance of the Hamiltonian, and apply the formula,
\begin{eqnarray}
\label{eq_integrate_inv}
\mbox{\boldmath{$C$}}(t)&=&\mbox{\boldmath{$C$}}(t_0)+\sum_{n=1}^{\infty}(-i)^n\frac{1}{n!}\int_{t_0}^t\!\! dt_{n-1}\int_{t_0}^{t} \!\! dt_{n-2}\! \cdots \!\!\int_{t_0}^{t}\!\!dt_0 \mbox{\boldmath{$\hat{H}$}}(t_{n-1})\mbox{\boldmath{$\hat{H}$}}(t_{n-2})\cdots \mbox{\boldmath{$\hat{H}$}}(t_{0})\mbox{\boldmath{$C$}}(t_0),\\
 &=& \sum_{n=0}^{\infty}\frac{(-i)^n}{n!}\left(\int_{t_0}^{t}\mbox{\boldmath{$\hat{H}$}}(t) dt\right)^n \mbox{\boldmath{$C$}}(t_0).
\end{eqnarray}
Using eq.(\ref{eq_h_deco}), we can easily derive a tentative expression of the probability amplitude from $J=r$ to $J=s$, given by
\begin{eqnarray}
\label{eq_solution}
C_{rs}(t) &=& \sum_{n=0}^{\infty} \sum_{k=0}^{N}\frac{1}{n!}\left( -\frac{i\gamma B_{\mbox{\scriptsize{f}}}}{2}\right)^n \left( \int_{t_0}^{t} \exp (\beta_s it)\cdot \exp(-\beta_r it) dt\right)^n y_{sk} \lambda_k^n y_{rk},\\
 &=& \sum_{n=0}^{\infty} \sum_{k=0}^{N}\left(\frac{\gamma B_{\mbox{\scriptsize{f}}}}{2(\beta_s-\beta_r)}\right)^n \frac{1}{n!} \left(\exp \{(\beta_s-\beta_r) it\} \right)^n y_{sk} \lambda_k^n y_{rk}.
\end{eqnarray}
An interesting characteristic of this expression is that it converges to the mathematically exact solution under a fully resonant condition as
\begin{equation}
\label{eq_limit}
\lim_{\mit\Delta B,D_v \to 0}\!\!C_{rs}(t) = \sum_{n=0}^{\infty} \sum_{k=0}^{N}\frac{1}{n!}\left(\frac{i\gamma B_{\mbox{\scriptsize{f}}} t \lambda_k}{2}\right)^n y_{sk} y_{rk}.
\end{equation}
The expression in eq.(\ref{eq_limit}) is regarded as the Bessel function if the network boundaries are located far from the initial state.
The expression derived here takes an extended form of the exact solution under a fully resonant condition. However, the role of $\gamma$ is changed between the fully resonant condition and others.
In eq.(\ref{eq_limit}), $\gamma$ and $t$ can be swapped, and the role of $\gamma$ is regarded as the velocity of wave packet propagation. However, in eq.(\ref{eq_solution}), $\gamma$ and $t$ cannot be swapped.
$\gamma$ can be considered to relate directly to the probability amplitude of each state. 

\subsection{Unified parameter}
The expression obtained in eq.(\ref{eq_solution}) cannot be used for calculating time-evolution because of its poor convergence property.
However, it seems natural that a term in the series might have information relating to the amplitudes.
We assume the expression below as representing the unified localization parameter, that is,
\begin{equation}
\label{eq_u_l}
u_{\mbox{\scriptsize{L}}}=\frac{\gamma \cdot B_{\mbox{\scriptsize{f}}}}{|\beta_s-\beta_r|}.
\end{equation}
The form of this parameter, particularly the dependence of $u_{\mbox{\scriptsize{L}}}$ on $\gamma$, $\beta_r$, and $\beta_s$ agrees with our intuition for localization.
This parameter is actually very useful for predicting the range of localization.
By comparing with numerical simulations, we obtain an empirical rule for localization using $u_{\mbox{\scriptsize{L}}}$.
The localization is found to occur in the region
\begin{equation}
\label{eq_u_l_half}
u_{\mbox{\scriptsize{L}}}>0.5.
\end{equation}
Outside this region, the probability amplitude decreases very rapidly.
The details of validation and discussion will be given in the remainder of this article.

\section{Numerical experiment}
Here we prove the validity of the value of obtained $u_{\mbox{\scriptsize{L}}}$ by comparison with numerical simulations.
We mainly solve eq.(\ref{eq_sch_ra}) using the fourth-order Runge-Kutta method.
Furthermore, we solve the set of eq.(\ref{eq_sch}) and eq.(\ref{eq_e_field}) when evaluating the validity of the rotating-wave approximation.
We perform a numerical demonstration of the isotope-selective rotational excitation between $^{39}$K$^{37}$Cl and $^{39}$K$^{35}$Cl as an example.
For the molecular constants, $D_v/B_M=8.21\times 10^{-7}$ for $^{39}$K$^{37}$Cl, and $8.45\times 10^{-7}$ for $^{39}$K$^{35}$Cl are used \cite{KCl}. 
The ratio $B_M$ of $^{39}$K$^{35}$Cl to that of $^{39}$K$^{37}$Cl is $1.03$.
The interval of pulses is tuned to synchronize the rotational period of $^{39}$K$^{37}$Cl.
The initial rotational state is set to $J=5$ to clearly show the localization with its upper and lower boundaries.
The results are given as relative values of a time-averaged probability distribution for each rotational state after irradiation with a great many pulses.
The maximum probability is always obtained for the initial state ($J=5$),
which is used as the standard for normalization.

The number of pulses for averaging is set to $\frac{1000}{\gamma}$ by considering the pulse of $\gamma =1$ increments $J$ by $1$; the increment is proportional to $\gamma$.
Under the rotating-wave approximation, the number of calculation points between the two pulses is set to $10\cdot \gamma$ for a resonant isotope or $20\cdot \gamma$ for a non-resonant isotope. 
In the case without the rotating-wave approximation, the number of calculation points is set to $4000 \cdot \gamma$ or $8000 \cdot \gamma$ because the change in the electric field is steeper than that in the case with the rotating-wave approximation.
The numerical errors in the sum of the rotational distribution are less than $10^{-6}$ in the case with the rotating-wave approximation, or less than $10^{-3}$ in the case without the rotating-wave approximation.

\section{Results}
\subsection{Dependence on $\gamma$}
\begin{figure}[htbp]
  \includegraphics[keepaspectratio,height=6.0cm]{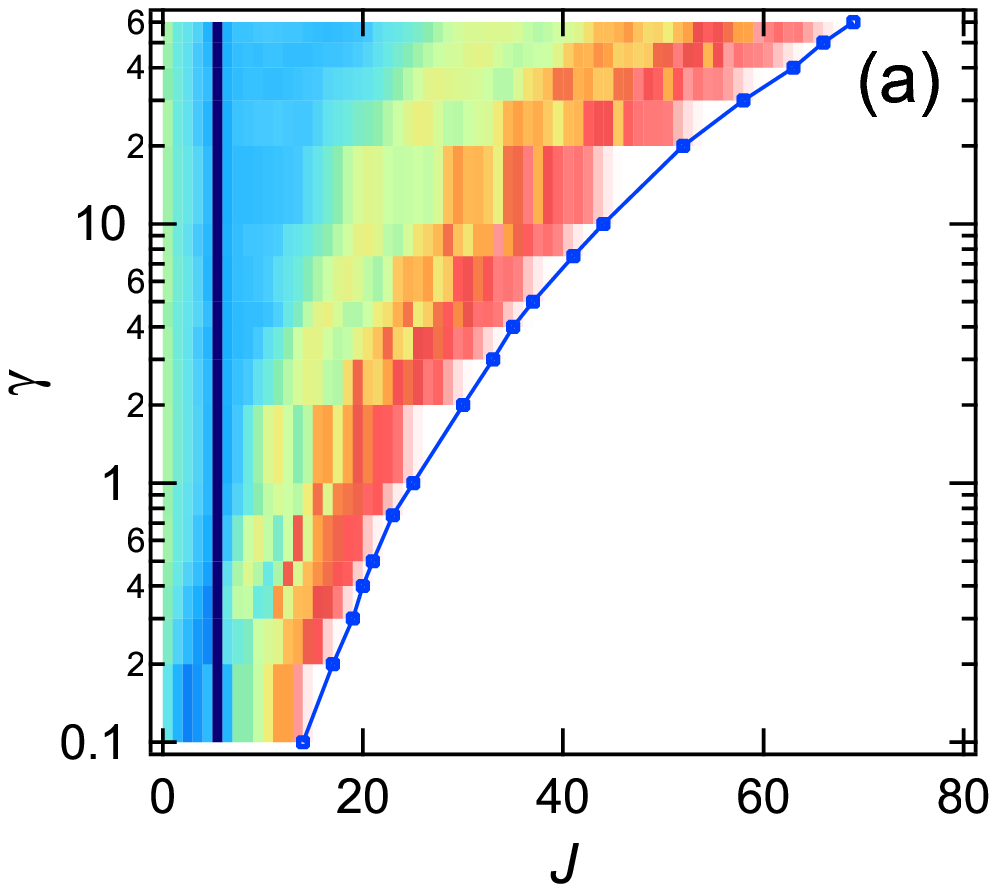}
  \includegraphics[keepaspectratio,height=6.0cm]{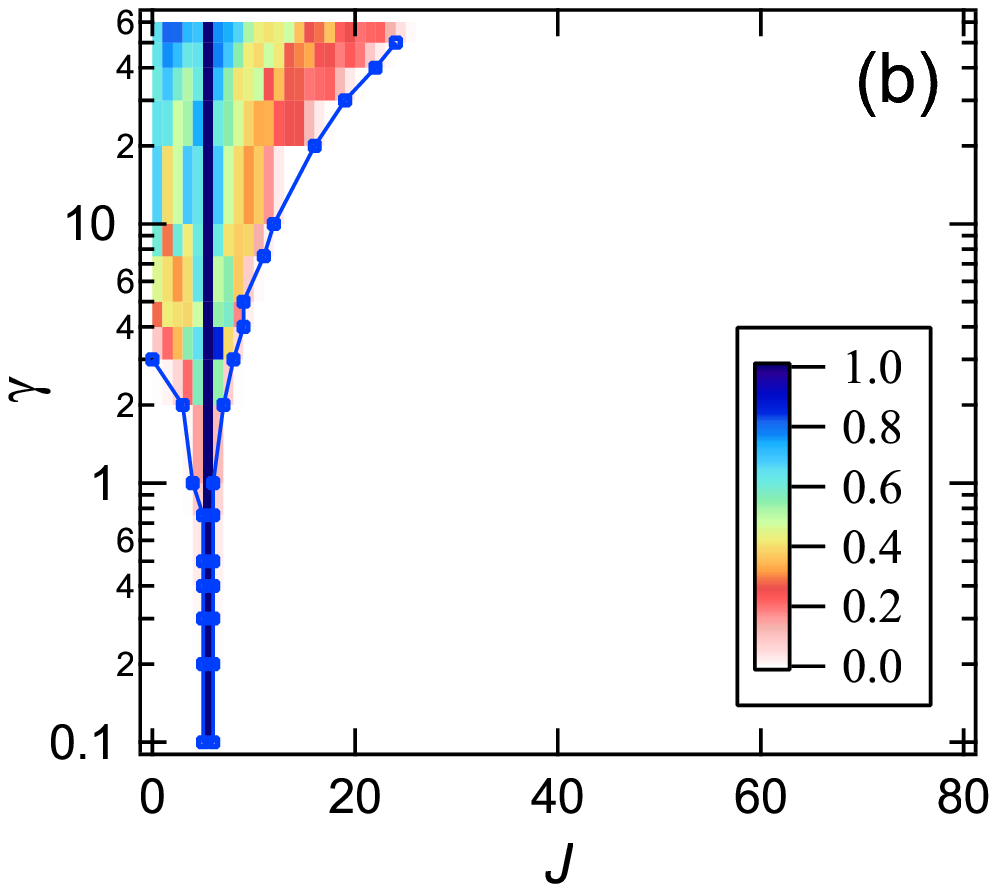}
  \caption{\label{fig_gamma}Time-averaged relative probability distribution with various $\gamma$ values calculated using eq.(\ref{eq_sch_ra}) and the boundary line of $u_L=0.5$ of $^{39}$K$^{37}$Cl (a) and $^{39}$K$^{35}$Cl (b).
The interval of pulses was synchronized with $T_{\mbox{\scriptsize{M}}}$ for $^{39}$K$^{37}$Cl.
The color scale is the same in both (a) and (b).}
\end{figure}

Figure \ref{fig_gamma} shows the results for the time-averaged probability distribution obtained with the numerical calculations using eq.(\ref{eq_sch_ra}) with various values of $\gamma$. 
The pulse interval $T_{\mbox{\scriptsize{p}}}$ was tuned to be $T_{\mbox{\scriptsize{M}}}$ of $^{39}$K$^{37}$Cl.
The boundaries of the time-averaged distribution were in complete agreement with the line $u_L=0.5$, and the dependence of the range on $\gamma$ was successfully explained.
With the rotating-wave approximation, eq.(\ref{eq_u_l}) was able to completely characterize the localization of the distribution.

\begin{figure}[htbp]
  \includegraphics[keepaspectratio,height=6.0cm]{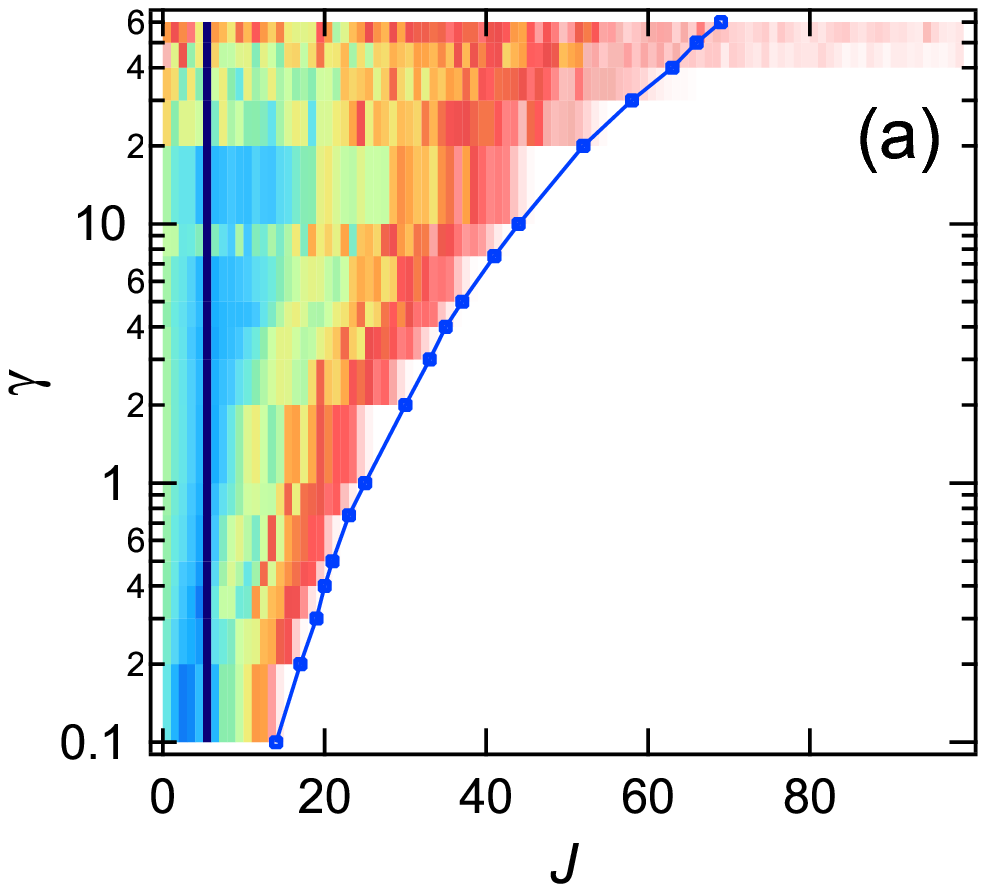}
  \includegraphics[keepaspectratio,height=6.0cm]{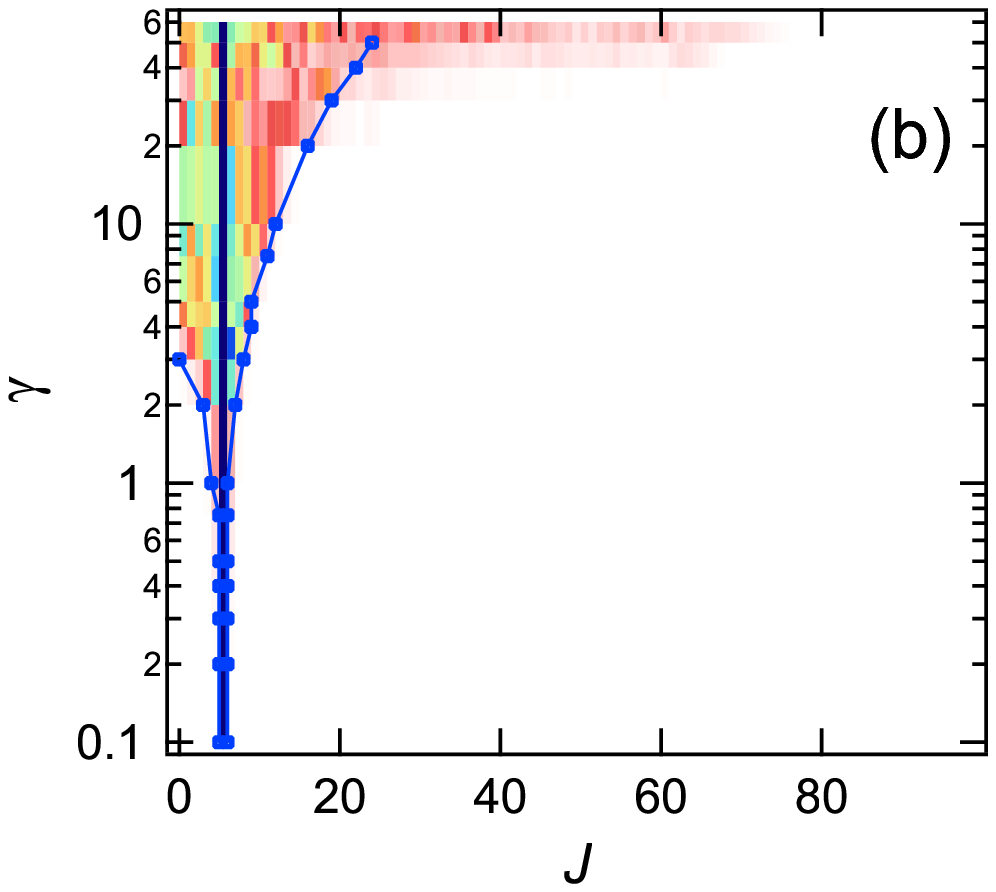}
  \caption{\label{fig_nwa}Same as Fig. \ref{fig_gamma} but calculated without the rotating-wave approximation.
The scale of color is the same as that for Fig. \ref{fig_gamma}.}
\end{figure}

The result of the numerical calculations without the rotating-wave approximation are shown in Fig. \ref{fig_nwa} for comparison. 
The calculated time-averaged distribution begins to show small differences from Fig. 3 around $\gamma=10$, and widely disperses at $\gamma > 40$.
The unified parameter $u_{\mbox{\scriptsize{L}}}$ was still useful in the region $\gamma \le 20$.

As shown from the results above, $u_{\mbox{\scriptsize{L}}}$ can be used to predict the upper and lower boundaries of the rotational distribution under the conditions of the rotating-wave approximation, regardless of variations in the pulse interval and centrifugal distortions. At higher $\gamma$ values, where the rotating-wave approximation cannot be applied, the probability distribution spreads over a wider range than predicted with $u_{\mbox{\scriptsize{L}}}$ in most cases.

\subsection{Dependence on $T_{\mbox{\scriptsize{p}}}$}
\begin{figure}[htbp]
  \includegraphics[keepaspectratio,height=6.0cm]{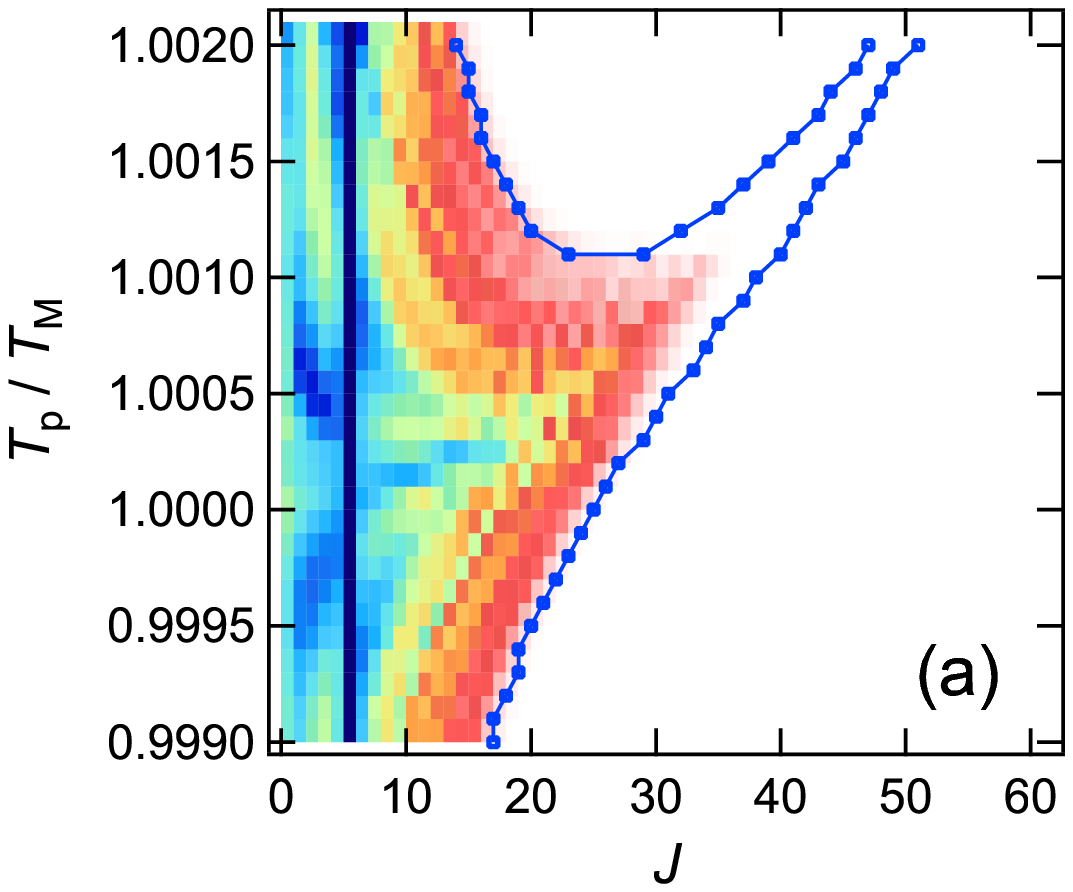}
  \includegraphics[keepaspectratio,height=6.0cm]{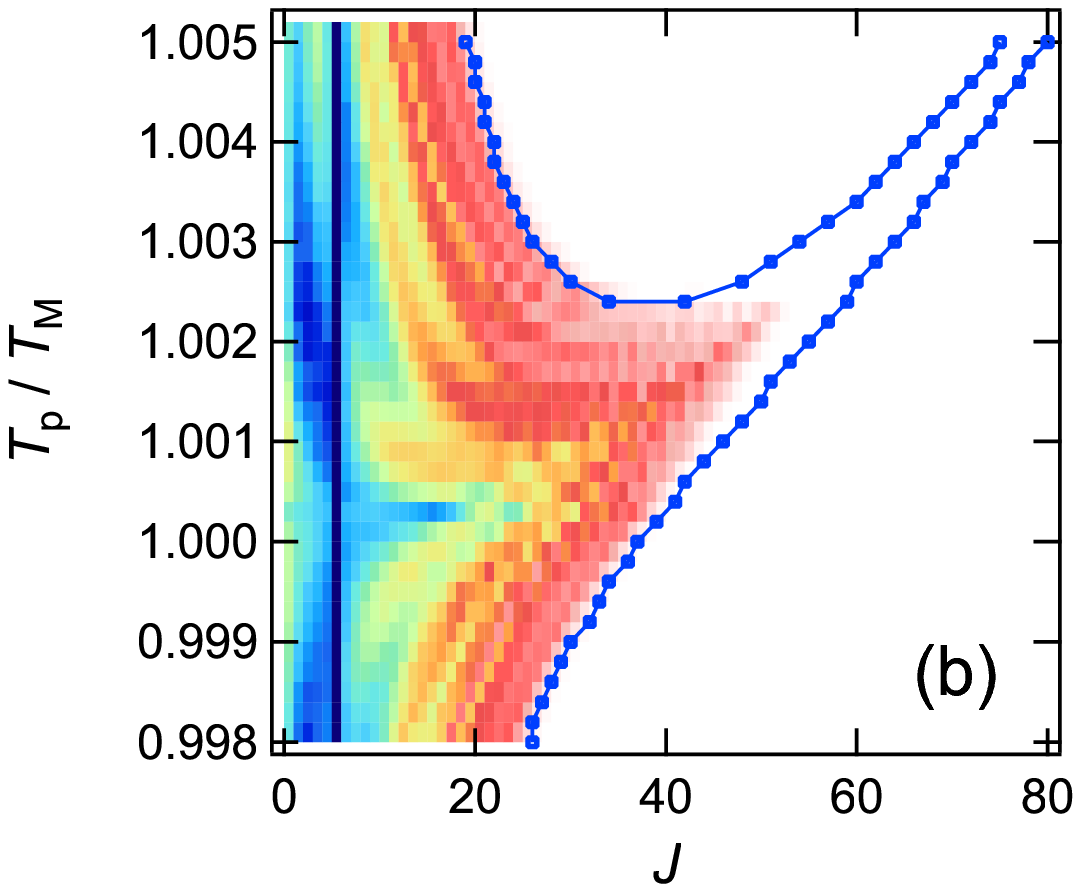}
  \caption{\label{fig_tp}Time-averaged relative probability distribution with various $T_{\mbox{\scriptsize{p}}}$ values calculated using eq.(\ref{eq_sch_ra}) and the boundary line of $u_{\mbox{\scriptsize{L}}}=0.5$ of $^{39}$K$^{37}$Cl  with $\gamma =1$(a) and $\gamma =5$(b).
The scale of color is the same as that in Fig. \ref{fig_gamma}.
Note that both the horizontal and vertical axes are different from each other in (a) and (b).}
\end{figure}

Next, we evaluated the validity of $u_{\mbox{\scriptsize{L}}}$ with various values of $T_{\mbox{\scriptsize{p}}}$ by fixing $\gamma$.
Figure \ref{fig_tp} shows the results of $\gamma=1$ and $\gamma=5$.
The rotating-wave approximation was applied because the values of $\gamma$ were sufficiently small.
By extending the pulse interval slightly from the rotational period of the molecules, the range of the time-averaged rotational distribution became slightly wider. 
This is because slight shifts by $D_v(J+1)^3$ were subtly compensated by tuning the peaks of the frequency comb. 
From Fig. \ref{fig_tp}(a) and (b), we can see that the best pulse interval for maximum compensation depends on $\gamma$. 
The $u_{\mbox{\scriptsize{L}}}$ successfully represents the limit of the compensation for each $\gamma$.

However, the $u_{\mbox{\scriptsize{L}}}$ boundary lines deviate from the calculated distribution at $T_{\mbox{\scriptsize{p}}}=1.0005-1.0010$ in (a), and $T_{\mbox{\scriptsize{p}}}=1.0010-1.0025$ in (b).
The intuitive reason for this is that the local minimum of $u_{\mbox{\scriptsize{L}}}$ prevented wave packet propagation.
In other words, for the frequency comb, the frequency gap between absorption and the comb becomes large in the middle region of $J$.

\section{Discussion}
First, we made a drastic assumption by ignoring the dependence of the interaction on quantum paths.
The analytical expression obtained could not be used for calculation of the exact time-evolution; however, its coefficient was observed to be a useful parameter.
The reason we could obtain approximate information from the coefficient probably relates to the characteristics of the spectral decomposition.
In the spectral decomposition, the Hamiltonian matrix, which depends on quantum paths, is expanded into repeating diagonal interactions with a basis transformation at the beginning and the end.
In the case without localization, the approximate time-evolution can be calculated using the same analytical method by ignoring the dependence on quantum paths \cite{chin}.
In the case with localization, we probably have to obtain a first-order approximation as a coefficient that can be related to the approximate probability amplitude.

There seems to be a presupposition that the unified parameter $u_{\mbox{\scriptsize{L}}}$ can be used only under conditions with the existence of quantum paths from the initial state to the final state.
A good counter-example to this presupposition is presented in Fig. 5.
In these calculations, there are isolated regions where $u_{\mbox{\scriptsize{L}}} > 0.5$ in the upper right in both figures.
If the value of $u_{\mbox{\scriptsize{L}}}$ alone determined the probability amplitude,
this amplitude would also be observed in the isolated region.
However, there is actually no quantum path to the isolated region from the initial state,
and no probability is observed in the isolated region.

A natural question to ask is why is the boundary value 0.5?
In our study, unfortunately, this value was obtained only through empirical investigations.
However, there will be a firm mathematical reason why that boundary value should be 0.5.

In this study, we assume a constant transition moment between the rotational states.
At $M=0$, this assumption has little influence on the evolution of the rotational distribution.
At low $M$, the unified parameter is still usable because the component of the parameter is not directly related to the matrix of the transition moment.
Almost the same mathematical treatment can be applied because the transition moment matrix (eq.\ref{eq_V}) can be always numerically diagonalized.
At high $M$, the validity of the unified parameter should be evaluated in each case because the change in the matrix (eq.\ref{eq_V}) can be regarded as an increase or decrease in $\gamma$.
It is likely that a more general theoretical treatment including the difference in $M$ can be found.

\section{Conclusion}
In this study, we derived a unified parameter to predict the range of localization of the rotational distribution of diatomic molecules in a train of optical pulses.
The validity of the obtained parameter was evaluated by comparison with numerical simulations.
Although the theoretical treatment was partly tentative, the obtained parameter can almost completely explain the localization induced by both interval mismatch and centrifugal distortion, known as Anderson localization and Bloch oscillation, respectively.
The odd dependence on the intensity of the pulses was also explained.
Because the unified parameter comprises molecular energies and simple experimental parameters, the parameter is useful for the planning of experiments and development.
In this study, we only deal with the resonant pulses in the terahertz region; however, our method can be extended to non-resonant pulses.
The parameter derived for non-resonant pulses will predict the position of $J_{\mbox{\scriptsize{R}}}$, which was mentioned in a previous work by Flo{\ss} {\it et al} \cite{floss03}.

\begin{acknowledgments}
We acknowledge the helpful discussion with Akira Ichihara
and Etsuo Segawa, and thank Keiichi Yokoyama for
providing feedback on the manuscript.
This work was supported by JSPS KAKENHI Grant No. 26420875.
\end{acknowledgments}

\end{document}